# Continual task learning in natural and artificial agents

Timo Flesch[1], Andrew Saxe[2] and Christopher Summerfield[1]

[1]Department of Experimental Psychology, University of Oxford, Oxford, UK.
[2]Gatsby Computational Neuroscience Unit & Sainsbury Wellcome Centre, UCL, London, UK.
Correspondence: {timo.flesch, christopher.summerfield}@psy.ox.ac.uk

**Abstract**

How do humans and other animals learn new tasks? A wave of brain recording studies has investigated how neural representations change during task learning, with a focus on how tasks can be acquired and coded in ways that minimise mutual interference. We review recent work that has explored the geometry and dimensionality of neural task representations in neocortex, and computational models that have exploited these findings to understand how the brain may partition knowledge between tasks. We discuss how ideas from machine learning, including those that combine supervised and unsupervised learning, are helping neuroscientists understand how natural tasks are learned and coded in biological brains.

**Highlights**

- Both natural and artificial agents face the challenge of learning in ways that support effective future behaviour
- This may be achieved by different learning regimes, associated with distinct dynamics, and differing dimensionality and geometry of neural task representations
- Where two different tasks are learned, neural codes for task-relevant information may be factorised in neocortex
- Combinations of supervised and unsupervised learning mechanisms may help partition task knowledge and avoid catastrophic interference

**Declaration**

Authors declare no competing interests

## 1. Natural tasks

In the natural world, humans and other animals behave in temporally structured ways that depend on environmental context. For example, many mammals cycle systematically through daily activities such as foraging, grooming, napping, and socialising. Humans live in complex societies in which labour is shared among group members, with each adult performing multiple successive roles, such as securing resources, caring for young, or exchanging social information. In many settings, we can describe the behaviour of natural agents as comprising a succession of distinct *tasks* for which a desired outcome (reward) is achieved by taking actions (responses) to observations (stimuli) through the learning of latent causal processes (rules).

The nature of task-driven behaviour, and the way that tasks are represented and implemented in neural circuits, have been widely studied by cognitive and neural scientists. One important finding is that switching between distinct tasks incurs a cost in decision accuracy and latency [1]. This *switch cost* implies the existence of control mechanisms that ensure we remain "on task", possibly protecting ongoing behavioural routines from interference [2,3]. In primates, there is good evidence that control signals originate in the prefrontal cortex (PFC) and encourage task-appropriate behaviours by biasing neural activity in sensory and motor regions [4]. For example, single cells in the PFC have been observed to respond to task rules [5], and patients with PFC damage tend to select tasks erroneously, leading to disinhibited or inappropriate behaviours [6].

How, then, are tasks coded in the PFC and interconnected regions? One key insight is that mutual interference among tasks can be mitigated when they are coded in *independent subspaces* of neural activity, such that the neural population vector evoked during task A is uncorrelated with that occurring during task B [7,8]. Over the past decade, evidence for this coding principle has emerged in domains as varied as skilled motor control [9], auditory prediction [10], memory processes [11,12], and visual categorisation [13,14]. However, the precise computational mechanisms by which tasks are encoded and implemented remain a matter of ongoing debate.

## 2. Rich and lazy learning

One way to study how tasks could be neurally encoded is to simulate learning in a simple class of computational model – a neural network trained with gradient descent. Neural networks uniquely allow researchers to form hypotheses about how neural codes form in biological brains, because their representations emerge through optimisation rather than being hand-crafted by the researcher [15]. One recent observation is that neural networks can learn to perform tasks in different regimes that are characterised by qualitatively diverging learning dynamics and distinct neural patterns at convergence [16,17]. In the *lazy regime*, which occurs when network weights are initialised with a broader range of values (e.g., higher connection strengths), the dimensionality of the input signals is rapidly expanded via random projections to the hidden layer such that learning is mostly confined to the readout weights, and error decreases exponentially [17–20]. By contrast, in the *rich regime*, which occurs when weights are initialised with low variance (weak connectivity), the hidden units learn highly structured representations that are tailored to the specific demands of the task, and the loss curve tends to pass through one or more saddle points before convergence [21–24]. We illustrate using a simple example – learning an "exclusive or" (XOR) problem – in **Fig. 1A-D**.

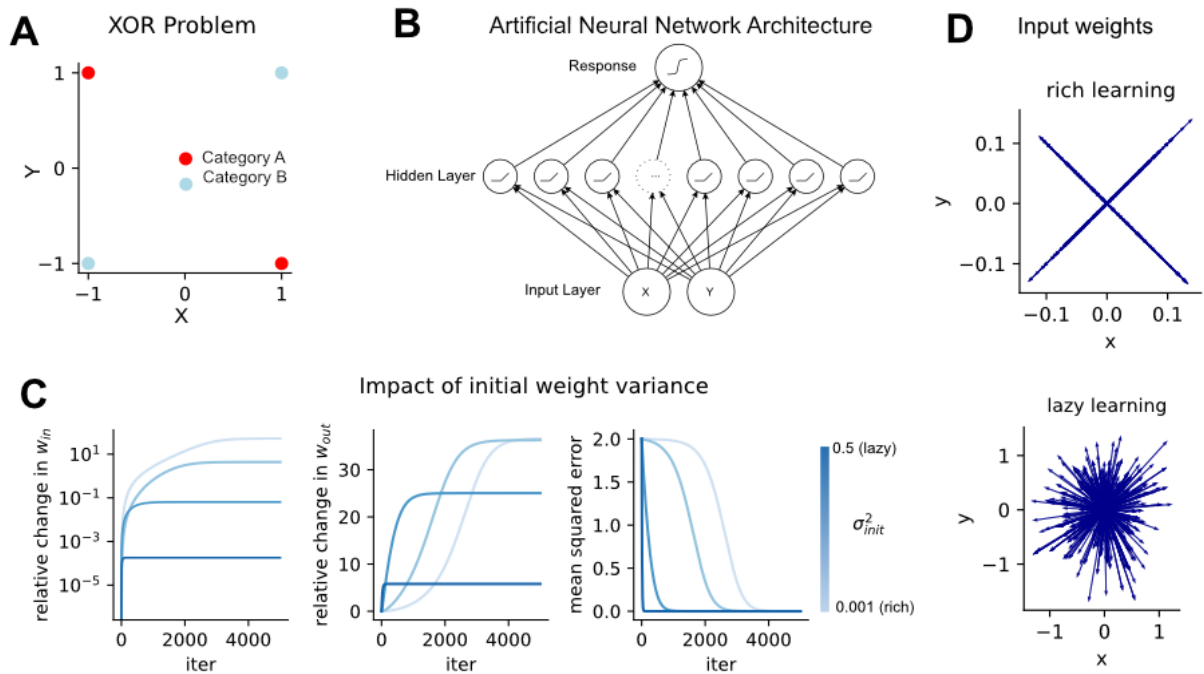

**Figure 1. Rich and Lazy learning in neural networks.** (**A**) The XOR (exclusive or) problem requires to provide the same response A when either one or the other of two input units are set to one, and the response B when both are 0 or both are 1. A linear classifier can't learn to distinguish between the two classes. (**B**) Feedforward neural network architecture that can solve the XOR task. The inputs are mapped into a hidden layer with non-linear outputs, and from there into a non-linear response layer. (**C**) Effect of different initial weight variances on the training dynamics of the network shown in (B). We distinguish between rich learning (with small initial weight variance, light blue) and lazy learning (with large initial weight variance, dark blue). The change of the magnitude of the input-to hidden (left) and hidden-to output (middle) depends strongly on initialisation strength. In lazy initialised networks, the input-to hidden weights remain very close to their initial values and learning is confined to the readout weights. In rich initialised networks, all weights adapt substantially. Moreover, rich-initialised networks learn much slower than lazy-initialised networks (right). (**D**) Learned input-to-hidden weights after rich (top) and lazy (bottom) learning. Under rich learning, the weights learn to point to the four input types. In contrast, under lazy learning, the weights point in arbitrary directions, effectively performing a random mapping into a higher-dimensional space.

These schemes may have complementary costs and benefits. High-dimensional coding schemes maximise the number of discriminations that can be linearly read out from the network, allowing agents to rapidly learn a new decision rule for a task [25]. Low-dimensional coding schemes confer robustness through redundancy, because neurons exhibit overlapping tuning properties, and promote generalisation, because they tend to correspond to simpler input-output functions when the neural manifold extends in fewer directions [26,27].

Neural recordings have offered evidence for both rich and lazy coding schemes. One important observation is that the variables that define a task – observations, actions, outcomes and rules – are often encoded jointly by single neurons. For example, when monkeys make choices on the basis of distinct cues, single cells tend to multiplex input and choice variables [28,29]. In another study, the dimensionality of neural codes recorded during performance of a dual memory task was found to approach its theoretical maximum, implying that neurons represent every possible combination of relevant variables [12]. This finding is consistent with "lazy" learning, implying that brains encode tasks via high-dimensional representations that enmesh multiple task-relevant variables across the neural population.

However, there is also important evidence that neural systems learn representations that mirror the structure of the task, as might be predicted in the "rich" regime. For example, it is often observed that neurons active in one task are silent during another, and vice versa. For example, when macaques were trained to categorise morphed animal images according to two independent classification schemes, 29% of PFC neurons became active during a single scheme, whereas only 2% of neurons were active during both [30]. Much more recently, a similar finding was reported using modern two-photon imaging methods in the parietal cortex of mice trained to perform both a grating discrimination task and a T-maze task. Over half of the recorded neurons were active in at least one task, but a much smaller fraction was active in both tasks [31]. In other words, the brain learns to partition task knowledge across independent sets of neurons.

One recent study used a neural network model to explicitly compare the predictions of the rich and lazy learning schemes to neural signals recorded from the human brain [13]. They developed a task (similar to ref [30]) that involved discriminating naturalistic images in two independent contexts. Human participants learn to make "plant / don't plant" decisions about quasi-naturalistic images of trees with continuously varying branch density and leaf density, whose growth success was determined by leaf density in one "garden" (task A) and branch density in the other (task B) (**Fig. 2A**). Neural networks could be trained to perform a stylised version of this task under either rich or lazy learning schemes by varying the different initial connection strengths (**Fig. 2B**). Multivariate methods used to visualise the representational dissimilarity matrix (RDM) and corresponding neural geometry for the network hidden layer under either scheme revealed that they made quite different predictions (**Fig. 2C**). Under lazy learning, the network learned a high dimensional solution whose RDM simply recapitulated the organisation of the input signals (into a grid defined by "leafiness" and "branchiness"). This is expected because randomly expanding the dimensionality of the inputs does not distort their similarity structure. However, under the rich scheme, the network compressed information that was irrelevant dimension to each context, so that the hidden layer represented the relevant input dimensions (leafiness and branchiness) on two neural planes lying at right angles in neural state space (**Fig. 2C-D**). Strikingly, BOLD signals in the posterior parietal cortex (PPC) and dorsomedial prefrontal cortex (dmPFC) exhibited a similar dimensionality and RDMs revealed a comparable geometric arrangement onto "orthogonal planes", providing evidence in support of "rich" task representations in the human brain (**Fig. 2E-F**).

Neural systems can thus learn both structured, low-dimensional task representations, and unstructured, high-dimensional codes. In artificial neural networks, the emerging regime depends on the magnitude of initial connection strengths in the network [15]. In the brain, these regimes may arise through other mechanisms such as pressure toward metabolic efficiency (regularisation), or architectural constraints that enforce unlearned nonlinear expansions. Whilst it remains unclear when, how and why either coding scheme might be adopted in the biological brain, neural theory is emerging that may help clarify this issue. One recent paper explored how representational structure is shaped by specific task demands [32]. Comparing recurrent artificial neural networks trained on different cognitive tasks, the authors found that those tasks that required flexible input-output mappings, such as the context-dependent decision task outlined above, induced task-specific representations, similar to the ones observed under rich learning in feedforward networks. In contrast, for tasks that did not require such a flexible mapping, the authors observed completely random, unstructured representations. This suggests that representational geometry is not only determined by intrinsic factors such as initial connection strength, but flexibly adapts to the computational demands of specific tasks. Rich task-specific representations might therefore arise when there is a need to minimise interference between different tasks and perform flexible context-specific responses to the same stimuli.

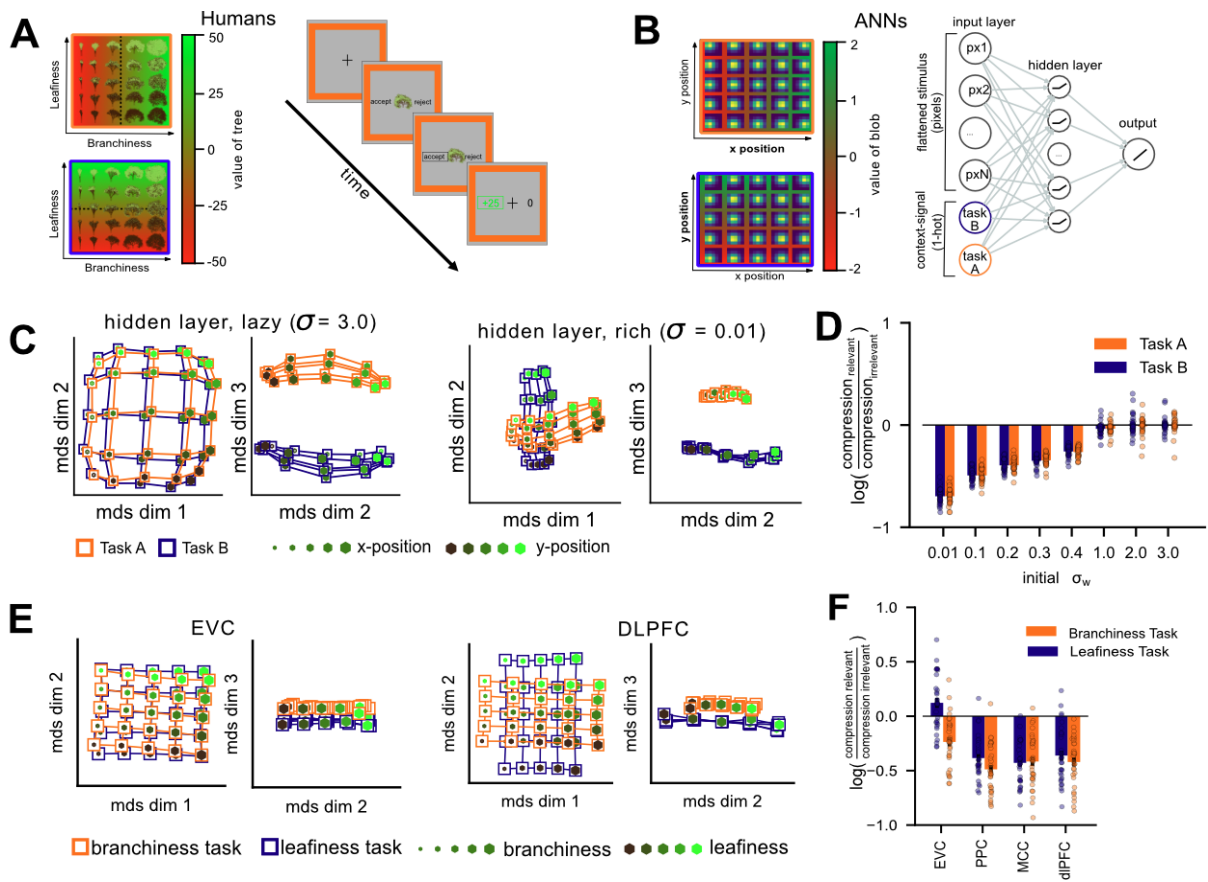

**Figure 2. Structured task representations for context-dependent decision making.** (**A**) Context-dependent decision making task with human participants. Stimuli were fractal images of trees that varied in their density of leaves (leafiness) and branches (branchiness). In each context/task only one of the two dimensions was relevant, indicated by a reward/penalty participants would receive for "accepting" the tree on a given trial. (**B**) Simplified version of task described in(A) with images of Gaussian "blobs" instead of trees. The mean of these blobs was varied in five steps along the x- and y-axis. In each context, only one of the two dimensions was relevant. A neural network (right) was trained to predict the context-specific feature value. (**C**) Hidden layer representations under lazy (left) and rich (right) learning. Under lazy learning, the network recapitulates the structure of the stimulus space. Under rich learning, it compresses along the task-irrelevant axes, forming "orthogonal" task-specific representations of relevant features. (**D**) Estimated compression rates as a function of the weight variance (rich vs lazy learning). Rich networks compress irrelevant dimensions more. (**E**) Visualisation of variance in human fMRI recordings from early visual cortex (EVC) and dorsolateral prefrontal cortex (DLPFC) explained by a model with free parameters for the compression rate, distance between tasks and rotation of individual task representations. Clearly evident are task-agnostic grid-like representations in EVC and "orthogonal" task-specific representational in DLPFC. (**F**) Estimated compression rates in different brain regions for human fMRI data, after training on the task described in (A).

## 3. The problem of continual learning

The natural world is structured so that different tasks tend to occur in succession. For example, many animals (such as bears and llamas) are able to both run and swim but they cannot do so at the same time. Similarly, most humans can perform many different tasks (such as playing the violin and the trumpet) but may not do so simultaneously. This aspect of the world presents a well-known challenge for task learning, because when task B is learned after task A, there is a risk that knowledge of task A is erased – a phenomenon known as "catastrophic forgetting" [33] (**Fig. 3A**). Catastrophic forgetting can occur when a neural network is optimised through adjustment of a finite set of connections, because there are no guarantees that any weight configuration that solves a novel task B will also simultaneously solve an existing task A (**Fig. 3B**). Building neural networks that avoid catastrophic forgetting and adapt continually to novel tasks in an open-ended environment is a grand challenge in AI research, where even powerful existing systems are often poor at adapting flexibly to new tasks that are introduced late in training [34–36]. Humans, by contrast, seem to have evolved ways to circumvent this problem, allowing people to solve new problems deep into older age. How might the neural representation of tasks allow biological agents to learn continually?

One possibility is that synapses that encode existing knowledge can be protected during new learning. Evidence for this idea comes from two-photon microscopy, which can be used to track dendritic spine formation in the rodent brain. When mice are trained to perform two unrelated tasks, such as running forward and backwards, new spines form across different apical tuft branches. Remarkably, many of these new spines are maintained stably across the entire lifespan of the animal, despite the many further experiences to which the animal is exposed, as if they were being protected from further change [37]. In machine learning, methods that earmark synapses that are critical for performing current tasks, and explicitly protect them during learning of new tasks, have helped neural networks learn several types of image recognition task or multiple Atari video games in succession [38,39].

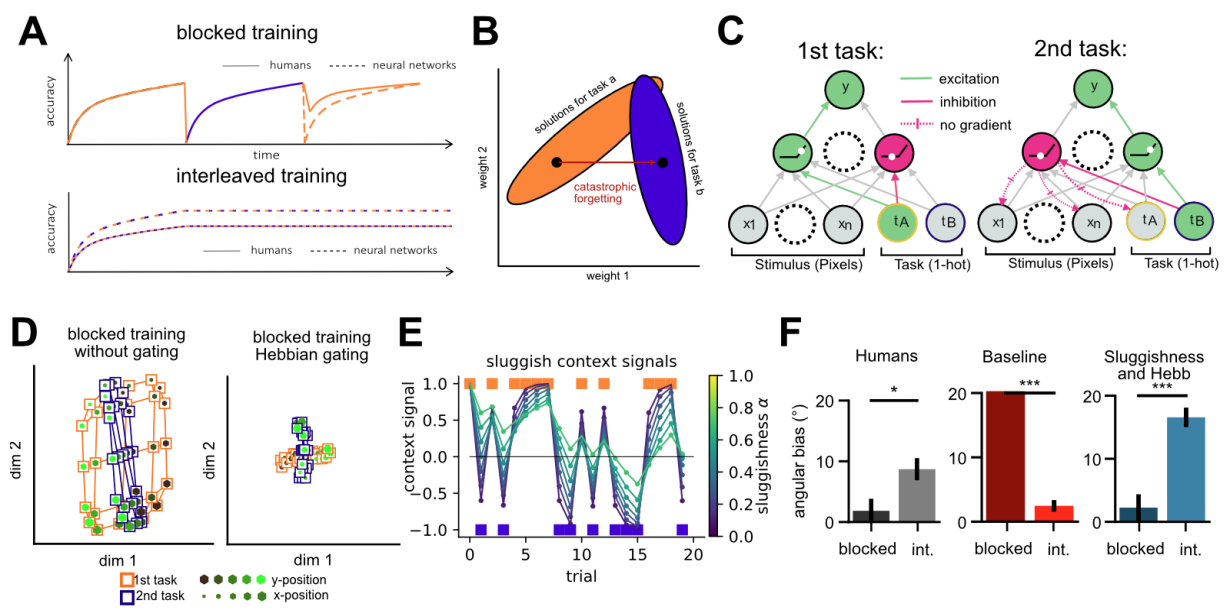

**Figure 3. Continual learning in minds and machines**. (A) Performance under blocked ("continual") and interleaved training in humans and deep neural networks. ANNs suffer from catastrophic forgetting under blocked training (top) but reach ceiling performance under interleaved training (bottom). The opposite is true for humans, wo perform worse under interleaved training. (B) Solution spaces in a neural network. Solutions for different tasks require different parameter configurations. Training on a second task moves the configuration out of the solution space for the first task. (C) Gating theory for continual learning. If the context signal was able to inhibit task-irrelevant units (which are relevant for another task), those should not be affected by gradient updates. (D) Hidden layer representations without (left) and with (right) Hebbian gating signals. The standard neural network treats the first task like the second. The gated network learns two separate task representations. (E) Sluggish context signals with exponentially weighted moving averages. The higher the "sluggishness", the more are context signals biased by the recent trial history. (F) Comparison of category boundary estimation error in humans (left) and a neural network either trained with standard error-corrective learning (mddle) or with Hebbian gating and sluggishness (right) after blocked (dark) and interleaved (light) learning. The cost of interleaved training is captured by larger estimation errors under interleaved learning.

Another promising approach to continual learning capitalises on the fact that in mammals, new information tends to be rapidly encoded in hippocampal circuits and only slowly consolidated to neocortex [40]. Consolidation allows existing learning (task A) to be internally "replayed" during acquisition of task B, removing the temporal structure from the learning problem by mentally interleaving tasks A and B. This allows the network to sample the two objectives simultaneously – and thus to gravitate towards a parameter setting that jointly solves both tasks [41]. This idea, known as "complementary learning systems" (or CLS) theory, influenced AI researchers designing the first deep neural networks to solve complex, dynamic control problems (such Atari games) [42]. By introducing a virtual "buffer" in which game states were stored and intermittently replayed, akin to the fast rehearsal of successive mental states during short wave ripples observed in both rodents [43] and humans [44], the network was able to overcome the challenge presented by a nonstationary training distribution, and to learn to perform dozens of games better than an expert human player [45]. More recently, new methods have been developed which prioritise replay of those states with highest prediction error [46], just as biological replay seems to privilege rewarding events [47], or that replay samples from a generative model of the environment, allowing agents to incorporate plausible but counterfactual events and outcomes into a joint task schema [48].

*4. The benefit of temporal structure*

Complementary learning systems theory offers a clear story about how the brain learns tasks in a temporally structured world – by mentally mixing them up. Indeed, there is evidence from cognitive psychology that injecting variety into the training set helps people learn sports [49], foreign languages [50] and even to acquire abstract mathematical concepts [51] or recognise the painting styles of famous artists [52]. However, as every student knows, simply selecting training examples at random rarely makes a good curriculum. When researchers want a novice animal to learn a complex task, they carefully design the series of training levels – for example by first teaching a mouse to lick a waterspout, and only later to discriminate a tone for reward. Pigeons struggle to learn the concept of "same" vs. "different" in visual discrimination if trained on random examples, but learn effectively if the training set starts small and grows gradually [53]; for related work in monkeys see [54]. Similarly, when training human children, teachers typically organise information into discrete blocks, so that for example students might study French before Spanish, but not both languages in the same lesson. Experimental animals may even spontaneously structure their own curricula, for example learning independently to lick a waterspout and discriminate a tone at different times [55]. In other words, although there are good theoretical reasons why mixing exemplars during training should help, in practice animals seem to learn more readily under temporally autocorrelated curricula.

Building on this intuition, one study used the trees task to ask whether curricula that block or interleave tasks (or "gardens", each associated with an independent visual discrimination rule) over trials facilitate learning [56]. Different groups of participants learned to perform the task by purely trial and error under either a *blocked* curriculum (involving hundreds of trials of consecutive practice for each task) or an *interleaved* curriculum (in which the task switched from trial to trial), before being tested without feedback on the interleaved task. Surprisingly, test performance was better for groups that learned under the blocked curriculum, despite the fact that other groups could in theory benefit from the shared structure of training and test. For example, the test involved many task switch trials, which the interleaved group had practiced extensively, but the blocked group had only experienced once. Detailed behavioural analysis showed that rather than increasing psychophysical sensitivity, or decreasing generalised errors (lapses), the blocked curriculum helped participants learn to apply the two category boundaries independently – in other words, it facilitated effective partitioning of the two tasks [56] (**Fig. 3F**).

Why does temporal structure assist learning in humans and other animals – but not in neural networks, who suffer catastrophic forgetting during blocked training? One possibility is that biological brains have evolved ways to partition knowledge during initial task acquisition, so that learning can occur in independent subspaces of neural population activity, thereby precluding interference between tasks A and B. Indeed, the orthogonal neural geometry of the BOLD signal observed for the trees task [13] occurred after *blocked* training – as if human participants, despite the temporal structure, had learned to represent the task in separate neural subspaces. As a proof of concept, methods that project learning updates (gradients) for new tasks into independent neural subspaces have been shown to help with continual learning in machine learning models, including recurrent networks [57,58]. However, implementing these methods often requires expensive computations, such as computing and maintaining an unwieldy projection matrix in memory.

To understand how the brain might partition new task learning across neurons, it is helpful to return to the simple neural network model of task learning. The widely-studied case where inputs such as trees [13,56], animals [30], or colour moving dots [29,59,60] are classified according to two independent category boundaries has XOR structure and thus can only be solved by networks equipped with nonlinear transformations, for example with rectified linear units (ReLU). When the task varies from trial to trial, the network is provided with "task inputs" denoting whether the current task is A or B (e.g., motion or colour discrimination). At initialisation, the weights connecting each of the task inputs to the hidden layer units are entirely random, but over the course of training they become anticorrelated, so that a hidden unit that responds positively to task A will tend to respond negatively to task B and vice versa (**Fig. 3C**). The additional rectification step – in which the ReLU sets the negative portion of an input to zero – means that hidden layer units become responsive to either one task input unit or the other, but not both, consistent with the finding that tasks tend to be partitioned across neurons, for example in mouse parietal cortex [31] or monkey PFC [30]. The partitioning allows tasks to be learned independently, and indeed if task input units weights are manually forced to be anticorrelated, the network has no problem learning both tasks A and B even when they are presented in successive blocks of training [61,62] (**Fig. 3D**). This is consistent with earlier proposals of context-based gating as a potential solution to continual learning and control [63–65].

*5. Knowledge partitioning via Hebbian learning*

The question remains, however, of how projection into partitioned hidden units might be achieved during online learning. A broad hint comes from considering the relationship between sensory input and the demands of natural tasks. In the real world, different tasks tend to occur in different contexts. For example, a bear might learn to walk on land and swim in water, and not vice versa. An adult in a professional role might work in the office, and drink the local bar, but not vice versa. Sensory context, thus – whether the backdrop is computers and desks, or beer and jukebox – offers strong clues about which tasks we should be performing [66]. Thus, a mechanism that learned to correlate task inputs that shared sensory features, and to orthogonalize those that did not, would allow knowledge to be partitioned by task. Fortunately, one popular learning algorithm neatly meets these requirements. An implementation of Hebbian learning called Oja's rule strengthens connections between neurons with covarying activity [67]. It thus converges to the first principal component of mean-centred input signals and will inevitably group together those hidden units that are connected to commonly activated inputs. For example, in a setting where tasks A and B occur independently, Oja's rule will tend to orthogonalize the weights from the two task input units to the hidden layer. This effect will be enhanced where tasks are accompanied by shared sensory features (such as land and water).

To capture the empirically observed advantage of blocked over interleaved learning, however, one further assumption is required – namely, that the window of temporal integration for the inputs is longer than a single trial. In other words, we assume that the input to the network on any given trial contains a mixture of the current and past information, so that decisions on any given trial may be biased by the recent trial history. Indeed sequential effects across trials are a ubiquitous feature of behavioural data gathered in the lab [68–70]. For context-dependent decisions, choice history will have a different effect where tasks are blocked and interleaved, because it effectively smooths the input signals so that independence between tasks (and thus partitioning) is promoted when tasks are blocked but decreased when they are interleaved. Intuitively, where task A and B are interleaved over trials, each input contains a mixture of signals from both tasks, reducing their independence (**Fig. 3E**). This may also disrupt performance, because the previous task may bias responses on the current trial. Bringing these ideas together, one recent modelling study showed that the introduction of history effects and Hebbian updating together allow networks to learn in ways that avoid catastrophic interference, and also to capture rich pattern of behaviour observed when human participants perform the trees task under blocked and interleaved conditions, including the advantage of blocking over interleaving, and the observation that this benefit stems from a more accurate estimate of the category boundary [61] (**Fig. 3F**).

*6. Learning tasks with and without supervision*

Machine learning models are often more likely to converge to optimal solutions when training data are sampled to be independent and identically distributed (i.i.d.) – in other words, with curricula that are as random as possible. Where the data distribution is stationary, i.i.d. sampling reduces bias during learning, and has powered machine learning models towards superhuman performance in domains such as object recognition [71]. However, the data distribution in the natural world is not stationary: instead, it is highly autocorrelated within a single context, but shifts abruptly at context boundaries, such as when a penguin emerges from the sea onto the ice or when you leave your warm house and head out into the wintry street. Some machine learning methods treat this structure as a nuisance, and have found clever ways to try and remove it [42]. However, the theory above suggests that the brain has instead evolved to capitalise on this structure. It proposes that by using unsupervised learning methods – such as Hebbian learning – the brain learns to group ongoing experience into contexts, and to partition neural resources to learn about each context independently. By orthogonalizing task signals, behavioural routines can be stored in ways that minimise interference.

This idea taps into a longstanding theme in machine learning research – namely, that unsupervised and error-driven learning can work together to help agents solve natural tasks. In fact, early successes in deep learning employed unsupervised pre-training methods to structure neural representations before supervised fine-tuning [72]. When deep convolutional networks are trained to solve the trees task from pixels alone, pre-training on the tree dataset using a beta-variational autoencoder (β-VAE) accelerates subsequent supervised learning. The β-VAE uses self-supervised methods to learn the latent factors in the data (i.e., leafiness and branchiness), thus structuring representations according to the two subsequent decision-relevant dimensions. In a similar vein, when human participants were first asked to arrange samples of trees by their similarity (without knowing the decision-relevant axes) those whose prior tendency was to organise by leafiness and branchiness received more benefit from blocked training [56]. In other words, learning the structure of the world can help both humans and neural networks organise information along task-relevant axes, and may by at the heart of biological solutions to continual learning.

However, there is an important caveat to this theory. When we encounter a novel task, we rarely want to ignore everything we know about other tasks – in fact, past task knowledge can help as well as hinder current task learning. For example, a chef learning a new recipe will benefit from past cooking experience, or a programmer learning Python will probably benefit from past proficiency in MATLAB [73]. Thus, we often want to share representations between tasks, but strict partitioning of task knowledge into orthogonal subspaces reduces the positive as well as the negative effects of this transfer. In fact, knowing how to learn new tasks in a way that negotiates the trade-off between interference (negative transfer) and generalisation (positive transfer) is a key unsolved challenge in both neuroscience and AI research[34,74].

Answers to this question are only beginning to emerge, but one possibility is that the brain is particularly adept at *factorising* existing task knowledge into reusable subcomponents, which can then be recomposed to tackle novel challenges [75]. For example, when making predictions about sequences of dots presented on a ring, participants seem to combine primitives involving rotation and symmetry [76]. Neural signals seem to code independently for task factors, such as the position and identity of an object in a sequence [77,78]. By coding reusable task factors in independent subspaces of neural activity, they can be recombined in novel ways – for example, if a chef has learned to knead dough when baking bread, and make a tomato passata when cooking spaghetti, these skills can be combined when making pizza for the first time.

One recent study showed that Hebbian learning may also contribute to compositional solutions to difficult transfer problems [79]. Human participants were trained to map coloured shapes onto spatial responses made a mouse click, with each feature (e.g., colour) mapping onto a spatial dimension (e.g., the horizontal axis in Cartesian coordinates, or radial axis in polar coordinates). Critically, they were trained with feedback on a single exemplar from each dimension (e.g., all red shapes and all coloured squares) and then asked to make inferences about the location associated with novel objects (e.g., blue triangles). As in the trees task, performance was improved when training of each dimension was blocked (e.g., all red items preceded all squares). Neural networks learned to perform perfectly on training trials but failed to transfer, unless they were equipped with a Hebbian learning step that helped them learn independently about colour and shape. With the combination of Hebbian and supervised learning, networks learned to perform the task in ways that closely resembled humans [79].

*7. Concluding remarks*

A renewed interest in connectionist models as theories of brain function [15], and the advent of high-throughput recording and multivariate analysis methods [27] have collectively reinvigorated research into task learning and its neural substrates. However, exactly how (and to what extent) neural representations form as biological agents learn new tasks remains a mystery. Some theories propose that task-related neurons are supremely adaptive, especially in

PFC, implying that new tasks automatically beget new geometries of representation [80]. Indeed, neural codes measured in BOLD signals have been shown to adjust rapidly when participants are taught new relations among objects or positions, and this occurs both in the medial temporal lobe and frontoparietal network [81–84]. There is even one report that orientation selectivity in V1 can adjust as people learn to classify gratings over just a few hours of practice, as if the basic building blocks of vision were themselves quite labile [85]. However, neural signals recorded from experimental animals seem to change much more gradually with learning, and it is unclear if the latter is a quirk of humans – or perhaps of BOLD signals. Understanding exactly how representations change in both hippocampus and neocortex during new task learning is currently a major outstanding challenge for 21$^{st}$ century neuroscience. See *Outstanding Questions* box for further unresolved issues.

**Outstanding questions**

- Past learning can interfere with current task performance, but at other times it can be beneficial. How does the brain code for tasks in a way that trades off the costs and benefits of negative and positive transfer?

- Given that neural circuits exhibit experience-dependent plasticity during learning, how can old learning be preserved?

- What are the respective roles of different brain regions, including the hippocampus and neocortex, in facilitating continual learning?

**REFERENCES**

1.  Monsell S. Task switching. Trends Cogn Sci. 2003/03/18 ed. 2003;7: 134–140. doi:10.1016/s1364-6613(03)00028-7

2.  Botvinick MM, Braver TS, Barch DM, Carter CS, Cohen JD. Conflict monitoring and cognitive control. Psychol Rev. 2001/08/08 ed. 2001;108: 624–52.

3.  Badre D. On task: how our brain gets things done. Princeton: Princeton University Press; 2020.

4.  Miller EK, Cohen JD. An integrative theory of prefrontal cortex function. Annu Rev Neurosci. 2001/04/03 ed. 2001;24: 167–202. doi:10.1146/annurev.neuro.24.1.167

5.  Freedman DJ, Assad JA. Neuronal Mechanisms of Visual Categorization: An Abstract View on Decision Making. Annu Rev Neurosci. 2016;39: 129–47. doi:10.1146/annurev-neuro-071714-033919

6.  Shallice T, Burgess PW. Deficits in strategy application following frontal lobe damage in man. Brain. 1991;114 ( Pt 2): 727–41.

7.  Lewandowsky S, Li S-C. Catastrophic interference in neural networks. Interference and Inhibition in Cognition. Elsevier; 1995. pp. 329–361. doi:10.1016/B978-012208930-5/50011-8

8.  Willshaw DJ, Buneman OP, Longuet-Higgins HC. Non-Holographic Associative Memory. Nature. 1969;222: 960–962. doi:10.1038/222960a0

9.  Kaufman MT, Churchland MM, Ryu SI, Shenoy KV. Cortical activity in the null space: permitting preparation without movement. Nat Neurosci. 2014;17: 440–448. doi:10.1038/nn.3643

10. Libby A, Buschman TJ. Rotational dynamics reduce interference between sensory and memory representations. Nat Neurosci. 2021;24: 715–726. doi:10.1038/s41593-021-00821-9

11. Xie Y, Hu P, Li J, Chen J, Song W, Wang X-J, et al. Geometry of sequence working memory in macaque prefrontal cortex. Science. 2022;375: 632–639. doi:10.1126/science.abm0204

12. Rigotti M, Barak O, Warden MR, Wang XJ, Daw ND, Miller EK, et al. The importance of mixed selectivity in complex cognitive tasks. Nature. 2013;497: 585–90. doi:10.1038/nature12160

13. Flesch T, Juechems K, Dumbalska T, Saxe A, Summerfield C. Orthogonal representations for robust context-dependent task performance in brains and neural networks. Neuron. 2022; S0896627322000058. doi:10.1016/j.neuron.2022.01.005

14. Failor SW, Carandini M, Harris KD. Learning orthogonalizes visual cortical population codes. Neuroscience; 2021 May. doi:10.1101/2021.05.23.445338

15. Saxe A, Nelli S, Summerfield C. If deep learning is the answer, what is the question? Nat Rev Neurosci. 2021;22: 55–67. doi:10.1038/s41583-020-00395-8

16. Woodworth B, Gunasekar S, Lee JD, Moroshko E, Savarese P, Golan I, et al. Kernel and Rich Regimes in Overparametrized Models. arXiv:200209277 [cs, stat]. 2020 [cited 17 Jan 2021]. Available: http://arxiv.org/abs/2002.09277

17. Chizat L, Oyallon E, Bach F. On Lazy Training in Differentiable Programming. NeurIPS. 2018. Available: http://arxiv.org/abs/1812.07956

18. Jacot A, Gabriel F, Hongler C. Neural tangent kernel: Convergence and generalization in neural networks. 2018. pp. 8571–8580.

19. Arora S, Ge R, Neyshabur B, Zhang Y. Stronger generalization bounds for deep nets via a compression approach. 2018. Available: http://arxiv.org/abs/1802.05296

20. Lee J, Xiao L, Schoenholz SS, Bahri Y, Novak R, Sohl-Dickstein J, et al. Wide Neural Networks of Any Depth Evolve as Linear Models Under Gradient Descent. arXiv. 2019. Available: http://arxiv.org/abs/1902.06720

21. Saxe AM, McClelland JL, Ganguli S. A mathematical theory of semantic development in deep neural networks. Proc Natl Acad Sci U S A. 2019;116: 11537–11546. doi:10.1073/pnas.1820226116

22. Geiger M, Jacot A, Spigler S, Gabriel F, Sagun L, d'Ascoli S, et al. Scaling description of generalization with number of parameters in deep learning. J Stat Mech. 2020;2020: 023401. doi:10.1088/1742-5468/ab633c

23. Paccolat J, Petrini L, Geiger M, Tyloo K, Wyart M. Geometric compression of invariant manifolds in neural nets. arXiv:200711471 [cs, stat]. 2021 [cited 22 Mar 2021]. Available: http://arxiv.org/abs/2007.11471

24. Saxe A, Sodhani S, Lewallen SJ. Neural Race Reduction: Dynamics of Abstraction in Gated Networks. Proceedings of the 39th International Conference on Machine Learning,. 2022. pp. 19287–19309.

25. Fusi S, Miller EK, Rigotti M. Why neurons mix: high dimensionality for higher cognition. Curr Opin Neurobiol. 2016;37: 66–74. doi:10.1016/j.conb.2016.01.010

26. Gao P, Trautmann E, Yu B, Santhanam G, Ryu S, Shenoy K, et al. A theory of multineuronal dimensionality, dynamics and measurement. BioRxiv. 2019. Available: https://doi.org/10.1101/214262

27. Gao P, Ganguli S. On simplicity and complexity in the brave new world of large-scale neuroscience. Curr Opin Neurobiol. 2015/05/02 ed. 2015;32: 148–55. doi:10.1016/j.conb.2015.04.003

28. Raposo D, Kaufman MT, Churchland AK. A category-free neural population supports evolving demands during decision-making. Nature neuroscience. 2014;17: 1784–92. doi:10.1038/nn.3865

29. Mante V, Sussillo D, Shenoy KV, Newsome WT. Context-dependent computation by recurrent dynamics in prefrontal cortex. Nature. 2013;503: 78–84. doi:10.1038/nature12742

30. Roy JE, Riesenhuber M, Poggio T, Miller EK. Prefrontal cortex activity during flexible categorization. J Neurosci. 2010;30: 8519–28. doi:10.1523/JNEUROSCI.4837-09.2010

31. Lee JJ, Krumin M, Harris KD, Carandini M. Task specificity in mouse parietal cortex. Neuron. 2022; S0896627322006626. doi:10.1016/j.neuron.2022.07.017

32. Dubreuil A, Valente A, Beiran M, Mastrogiuseppe F, Ostojic S. The role of population structure in computations through neural dynamics. Nat Neurosci. 2022;25: 783–794. doi:10.1038/s41593-022-01088-4

33. French RM. Catastrophic forgetting in connectionist networks. Trends Cogn Sci. 1999;3: 128–135. doi:10.1016/s1364-6613(99)01294-2

34. Parisi G, Kemker R, Part JL, Kanan C, Wermter S. Continual Lifelong Learning with Neural Networks: A Review. Neural Networks. 2019. doi:10.1016/j.neunet.2019.01.012

35. Hadsell R, Rao D, Rusu AA, Pascanu R. Embracing Change: Continual Learning in Deep Neural Networks. Trends in Cognitive Sciences. 2020;24: 1028–1040. doi:10.1016/j.tics.2020.09.004

36. Dohare S, Sutton RS, Mahmood AR. Continual Backprop: Stochastic Gradient Descent with Persistent Randomness. 2021 [cited 1 Oct 2022]. doi:10.48550/ARXIV.2108.06325

37. Yang G, Pan F, Gan W-B. Stably maintained dendritic spines are associated with lifelong memories. Nature. 2009;462: 920–924. doi:10.1038/nature08577

38. Kirkpatrick J, Pascanu R, Rabinowitz N, Veness J, Desjardins G, Rusu AA, et al. Overcoming catastrophic forgetting in neural networks. Proc Natl Acad Sci U S A. 2017;114: 3521–3526. doi:10.1073/pnas.1611835114

39. Zenke F, Poole B, Ganguli S. Continual Learning Through Synaptic Intelligence. arXiv:170304200. 2017.

40. Alvarez P, Squire LR. Memory consolidation and the medial temporal lobe: a simple network model. Proc Natl Acad Sci USA. 1994;91: 7041–7045. doi:10.1073/pnas.91.15.7041

41. McClelland JL, McNaughton BL, O'Reilly RC. Why there are complementary learning systems in the hippocampus and neocortex: insights from the successes and failures of connectionist models of learning and memory. Psychol Rev. 1995;102: 419–57.

42. Kumaran D, Hassabis D, McClelland JL. What Learning Systems do Intelligent Agents Need? Complementary Learning Systems Theory Updated. Trends Cogn Sci. 2016;20: 512–534. doi:10.1016/j.tics.2016.05.004

43. Foster DJ. Replay Comes of Age. Annu Rev Neurosci. 2017;40: 581–602. doi:10.1146/annurev-neuro-072116-031538

44. Vaz AP, Wittig JH, Inati SK, Zaghloul KA. Replay of cortical spiking sequences during human memory retrieval. Science. 2020;367: 1131–1134. doi:10.1126/science.aba0672

45. Mnih V, Kavukcuoglu K, Silver D, Rusu AA, Veness J, Bellemare MG, et al. Human-level control through deep reinforcement learning. Nature. 2015;518: 529–33. doi:10.1038/nature14236

46. Schaul T, Quan J, Antonoglou I, Silver D. Prioritized Experience Replay. arXiv:151105952 [cs]. 2016 [cited 24 Dec 2021]. Available: http://arxiv.org/abs/1511.05952

47. Ambrose RE, Pfeiffer BE, Foster DJ. Reverse Replay of Hippocampal Place Cells Is Uniquely Modulated by Changing Reward. Neuron. 2016;91: 1124–1136. doi:10.1016/j.neuron.2016.07.047

48. van de Ven GM, Tolias AS. Generative replay with feedback connections as a general strategy for continual learning. arXiv:180910635 [cs, stat]. 2019 [cited 25 Sep 2020]. Available: http://arxiv.org/abs/1809.10635

49. Goode S, Magill RA. Contextual Interference Effects in Learning Three Badminton Serves. Research Quarterly for Exercise and Sport. 1986;57: 308–314. doi:10.1080/02701367.1986.10608091

50. Richland LE, Bjork, R.A. Differentiating the Contextual Interference Effect from the Spacing Effect. 2004.

51. Rohrer D, Dedrick RF, Stershic S. Interleaved practice improves mathematics learning. Journal of Educational Psychology. 2015;107: 900–908. doi:10.1037/edu0000001

52. Kornell N, Bjork RA. Learning Concepts and Categories: Is Spacing the "Enemy of Induction"? Psychol Sci. 2008;19: 585–592. doi:10.1111/j.1467-9280.2008.02127.x

53. Katz JS, Wright AA. Same/different abstract-concept learning by pigeons. Journal of Experimental Psychology: Animal Behavior Processes. 2006;32: 80–86. doi:10.1037/0097-7403.32.1.80

54. Antzoulatos EG, Miller EK. Differences between Neural Activity in Prefrontal Cortex and Striatum during Learning of Novel Abstract Categories. Neuron. 2011;71: 243–9.

55. Kuchibhotla KV, Hindmarsh Sten T, Papadoyannis ES, Elnozahy S, Fogelson KA, Kumar R, et al. Dissociating task acquisition from expression during learning reveals latent knowledge. Nat Commun. 2019;10: 2151. doi:10.1038/s41467-019-10089-0

56. Flesch T, Balaguer J, Dekker R, Nili H, Summerfield C. Comparing continual task learning in minds and machines. Proc Natl Acad Sci U S A. 2018;115: E10313–E10322. doi:10.1073/pnas.1800755115

57. Zeng G, Chen Y, Cui B, Yu S. Continual learning of context-dependent processing in neural networks. Nature Machine Intelligence. 2019;1: 364–372.

58. Duncker L, Driscoll L, Shenoy KV, Sahani M, Sussillo D. Organizing recurrent network dynamics by task-computation to enable continual learning. 2020.

59. Takagi Y, Hunt L, Woolrich MW, Behrens EJ, Klein-Flugge MC. Projections of non-invasive human recordings into state space show unfolding of spontaneous and over-trained choice. BiorXiv. 2020. Available: https://doi.org/10.1101/2020.02.24.962290

60. Brincat SL, Siegel M, von Nicolai C, Miller EK. Gradual progression from sensory to task-related processing in cerebral cortex. Proc Natl Acad Sci USA. 2018;115. doi:10.1073/pnas.1717075115

61. Flesch T, Nagy DG, Saxe A, Summerfield C. Modelling continual learning in humans with Hebbian context gating and exponentially decaying task signals. 2022 [cited 15 Sep 2022]. doi:10.48550/ARXIV.2203.11560

62. Russin J, Zolfaghar M, Park SA, Boorman E, O'Reilly RC. A Neural Network Model of Continual Learning with Cognitive Control. arXiv:220204773 [cs, q-bio]. 2022 [cited 7 Mar 2022]. Available: http://arxiv.org/abs/2202.04773

63. Masse NY, Grant GD, Freedman DJ. Alleviating catastrophic forgetting using context-dependent gating and synaptic stabilization. Proc Natl Acad Sci U S A. 2018;115: E10467–E10475. doi:10.1073/pnas.1803839115

64. Tsuda B, Tye KM, Siegelmann HT, Sejnowski TJ. A modeling framework for adaptive lifelong learning with transfer and savings through gating in the prefrontal cortex. Proc Natl Acad Sci USA. 2020;117: 29872–29882. doi:10.1073/pnas.2009591117

65. Cohen JD, Dunbar K, McClelland JL. On the control of automatic processes: A parallel distributed processing account of the Stroop effect. Psychological Review. 1990;97: 332–361. doi:10.1037/0033-295X.97.3.332

66. Bar M. Visual objects in context. Nat Rev Neurosci. 2004/07/21 ed. 2004;5: 617–29. doi:10.1038/nrn1476

67. Oja E. Simplified neuron model as a principal component analyzer. J Math Biology. 1982;15: 267–273. doi:10.1007/BF00275687

68. Yu A, Cohen J. Sequential effects: superstition or rational behavior? In: Koller D, Schuurmans D, Bengio Y, Bottou L, editors. Advances in Neural Information Processing Systems,. Vancouver; 2009. pp. 1873–1880.

69. Cho RY, Nystrom LE, Brown ET, Jones AD, Braver TS, Holmes PJ, et al. Mechanisms underlying dependencies of performance on stimulus history in a two-alternative forced-choice task. Cognitive, Affective, & Behavioral Neuroscience. 2002;2: 283–299. doi:10.3758/CABN.2.4.283

70. Akaishi R, Umeda K, Nagase A, Sakai K. Autonomous Mechanism of Internal Choice Estimate Underlies Decision Inertia. Neuron. 2014;81: 195–206. doi:10.1016/j.neuron.2013.10.018

71. Krizhevsky A, Sutskever I, Hinton GE. ImageNet Classification with Deep ConvolutionalNeural Networks. Lake Tahoe, Nevada; 2012.

72. Bengio Y, Lamblin P, Popovici D, Larochelle H. Greedy layer-wise training of deep networks. 2006.

73. Sheahan H, Luyckx F, Nelli S, Teupe C, Summerfield C. Neural state space alignment for magnitude generalization in humans and recurrent networks. Neuron. 2021;109: 1214-1226.e8. doi:10.1016/j.neuron.2021.02.004

74. Musslick S, Saxe A, Hoskin AN, Reichman D, Cohen JD. On the Rational Boundedness of Cognitive Control: Shared Versus Separated Representations. PsyArXiv. 2020 [cited 17 Mar 2022]. Available: https://psyarxiv.com/jkhdf/

75. Behrens TEJ, Muller TH, Whittington JCR, Mark S, Baram AB, Stachenfeld KL, et al. What Is a Cognitive Map? Organizing Knowledge for Flexible Behavior. Neuron. 2018;100: 490–509. doi:10.1016/j.neuron.2018.10.002

76. Amalric M, Wang L, Pica P, Figueira S, Sigman M, Dehaene S. The language of geometry: Fast comprehension of geometrical primitives and rules in human adults and preschoolers. Gallistel R, editor. PLoS Comput Biol. 2017;13: e1005273. doi:10.1371/journal.pcbi.1005273

77. Liu Y, Dolan RJ, Kurth-Nelson Z, Behrens TEJ. Human Replay Spontaneously Reorganizes Experience. Cell. 2019;178: 640-652 e14. doi:10.1016/j.cell.2019.06.012

78. Al Roumi F, Marti S, Wang L, Amalric M, Dehaene S. Mental compression of spatial sequences in human working memory using numerical and geometrical primitives. Neuron. 2021;109: 2627-2639.e4. doi:10.1016/j.neuron.2021.06.009

79. Dekker R, Otto F, Summerfield C. Determinants of human compositional generalization. PsyArXiv; 2022 Mar. doi:10.31234/osf.io/qnpw6

80. Duncan J. An adaptive coding model of neural function in prefrontal cortex. Nature reviews Neuroscience. 2001;2: 820–9. doi:10.1038/35097575

81. Nelli S, Braun L, Dumbalska T, Saxe A, Summerfield C. Neural knowledge assembly in humans and deep networks. Neuroscience; 2021 Oct. doi:10.1101/2021.10.21.465374

82. Milivojevic B, Vicente-Grabovetsky A, Doeller CF. Insight reconfigures hippocampal-prefrontal memories. Curr Biol. 2015;25: 821–30. doi:10.1016/j.cub.2015.01.033

83. Morton NW, Schlichting ML, Preston AR. Representations of common event structure in medial temporal lobe and frontoparietal cortex support efficient inference. Proc Natl Acad Sci USA. 2020;117: 29338–29345. doi:10.1073/pnas.1912338117

84. Schapiro AC, Rogers TT, Cordova NI, Turk-Browne NB, Botvinick MM. Neural representations of events arise from temporal community structure. Nat Neurosci. 2013;16: 486–92. doi:10.1038/nn.3331

85. Ester EF, Sprague TC, Serences JT. Categorical Biases in Human Occipitoparietal Cortex. J Neurosci. 2020;40: 917–931. doi:10.1523/JNEUROSCI.2700-19.2019